\documentclass[conference]{IEEEtran}
\IEEEoverridecommandlockouts
\usepackage{cite}
\usepackage{amsmath,amssymb,amsfonts}
\usepackage{algorithmic}
\usepackage{graphicx}
\usepackage{textcomp}
\usepackage{xcolor}
\usepackage{hyperref}
\def\BibTeX{{\rm B\kern-.05em{\sc i\kern-.025em b}\kern-.08em
    T\kern-.1667em\lower.7ex\hbox{E}\kern-.125emX}}
\begin{document}

\title{Quantum Processing Unit (QPU) processing time Prediction with Machine Learning}

\author{
\IEEEauthorblockN{
Lucy Xing, %\IEEEauthorrefmark{1}, 
Sanjay Vishwakarma, %\IEEEauthorrefmark{1}, 
David Kremer, \\%\IEEEauthorrefmark{1}, 
Francisco Martín-Fernández, %\IEEEauthorrefmark{1}, 
Ismael Faro %\IEEEauthorrefmark{1}, 
and Juan Cruz-Benito}%\IEEEauthorrefmark{1}}
\IEEEauthorblockA{%\IEEEauthorrefmark{1}
IBM Quantum, IBM T.J. Watson Research Center, \\Yorktown Heights, NY, USA}}%\\

\maketitle

\begin{abstract}
This paper explores the application of machine learning (ML) techniques in predicting the QPU processing time of quantum jobs. By leveraging ML algorithms, this study introduces predictive models that are designed to enhance operational efficiency in quantum computing systems. Using a dataset of about 150,000 jobs that follow the IBM Quantum schema, we employ ML methods based on Gradient-Boosting (LightGBM) to predict the QPU processing times, incorporating data preprocessing methods to improve model accuracy. The results demonstrate the effectiveness of ML in forecasting quantum jobs. This improvement can have implications on improving resource management and scheduling within quantum computing frameworks. This research not only highlights the potential of ML in refining quantum job predictions but also sets a foundation for integrating AI-driven tools in advanced quantum computing operations.
\end{abstract}


\section{Introduction}

The nature of quantum computing becomes increasingly relevant in the core of data-center operations due to a paradigm shift in computational processing, prioritization, and execution. At the intersection of advanced quantum computing and intricate software architecture lies a complex, heterogeneous landscape of resources. In this context, the QPU plays a pivotal role. Each distinct component of the software and hardware stack influences how a quantum job can be executed, and designing efficient systems is a challenging task \cite{johnson2022qiskit}. Furthermore, as the quantum computing field advances, there is a growing need for systems that are more robust, responsive, resilient, or adaptable to various execution modes and user needs, among other requirements. To some extent, quantum computing systems and stacks are expected to broaden the scope of current software design and architectural definitions \cite{taylor2009software, murillo2024challenges, zhao2024unraveling, yue2023challenges, khan2023software} to accommodate the unique characteristics and challenges of this new computing paradigm. Among the features that quantum computing systems may have, we can identify the integration of intelligent capabilities. These new smart features enable systems to respond to various events, enhance the user experience, optimize the resources required for quantum computing jobs, prevent internal errors, and guarantee that users' quantum programs perform efficiently and effectively within the stack, in terms of both performance and efficiency. This current work fits in the area of new smart features that could help the improvement of existing quantum computing systems and stacks.

Traditional models used to estimate computational time and resource requirements fall short in quantum environments, where the job execution time of computational primitives may depend on information learned at runtime. The diversity and complexity of quantum job behavior, combined with the evolving landscape of quantum programming techniques and hardware advancements, require a more flexible and complex approach to predict it to improve other existing methods which may not be accurate \cite{ma2024understanding}. ML emerges as an useful technique in this scenario, where patterns and data analysis comes to navigate the quantum computing ecosystem. By harnessing ML, we can derive predictive insights that are both nuanced and dynamic, catering to the unique attributes of quantum jobs and ensuring that the software architecture remains robust and efficient in the face of quantum computing's challenges and opportunities.
In this paper, we specifically examine the distinct patterns generated by quantum jobs in the given environment to anticipate their behavior in advance. To demonstrate its potential, we present novel methods for predicting QPU processing time for quantum jobs. To clarify, QPU time refers to the duration that the QPU is locked to execute a particular quantum computing job. QPU time is a crucial factor in assessing the overall performance and efficiency of quantum computing systems.

This paper aims to highlight the path towards an AI-driven approach to job behavior prediction for quantum jobs, offering a comprehensive analysis that bridges the gap between its potential and practical application. The primary objective of this research is to illustrate the application of Artificial Intelligence (AI) in anticipating quantum job behavior, specifically in relation to QPU time. The secondary goal is to discuss how these features could enhance existing quantum computing systems by improving their performance, efficiency, and overall capabilities.

The paper is organized as follows: Section II presents the materials and methods used for this paper. Section III presents the main results related to predicting job's QPU time. Section IV discusses the results, presents some conclusions and future directions for this work.

\section{Materials and methods}

\subsection{Materials}

\subsubsection{Data Collection}

The dataset utilized in this study consists of anonymized data from genuine quantum jobs sourced from IBM Quantum's database. This dataset excludes any personal information or personal identifiable information. Quantum jobs data are accumulated daily, pre-processed and archived in pickle files for further analysis. The data used in this research was gathered between 2025-03-05 and 2025-03-21. In total, 166,143 quantum jobs were utilized for QPU time prediction. Each quantum job is recorded with 43 distinct data fields. These fields encompass elements such as the IBM Quantum backend the job will run on, the primitive id of the job (sampler/estimator), the total number of requested shots, and error suppression and correction methods, among others. Metadata such as the total number of executions and the number of batches are used as a proxy to determine the approximate job size and play a significant role in the prediction of QPU times.

To train the models, a configuration file is utilized to define the parameters\footnote{\href{https://lightgbm.readthedocs.io/en/latest/Parameters.html}{LightGBM parameters}} for the input features (X) and the output feature (Y). In this instance, the input features for X consist of the quantum job metadata, while the output feature for Y is the QPU time. The X column parameters are categorized into various data types, such as numerical and categorical, for encoding purposes, which will be elaborated in the Methods section.

\subsubsection{Model}

We trained a ML model to predict QPU time using the Light Gradient-Boosting Machine (LightGBM) library \cite{ke2017lightgbm}. LightGBM is an open-source gradient-boosting framework based on decision trees, and designed to be highly efficient. LightGBM uses histogram-based algorithms in combination with leaf-wise tree growth algorithms that has the advantage of faster training, being highly accurate, and reducing memory usage --- factors that are critical for datasets of larger scale\footnote{\href{https://lightgbm.readthedocs.io/en/stable/Features.html}{LightGBM features}}. In addition, LightGBM also allows for weights to be set on the training data. Since recent data more closely resemble the latest quantum job behavior in terms of QPU time taken, we assigned higher weights to more recent data and comparatively lower weights to older data when training our model.

Based on our feature importance analysis for predicting QPU execution time, presented below is a small subset of the most important features that our model uses:
\begin{itemize}
  \item backend – The QPU type significantly influences execution time due to hardware-specific latencies and queuing behavior.
  \item primitive\_id – Identifies a sampler or estimator primitive. 
  \item sum\_shots – The total number of shots strongly correlates with execution duration, as more measurements lead to longer QPU usage.
  \item sum\_durations\_per\_pub – Aggregated the duration of each Primitive Unified Bloc (PUB)\footnote{\href{https://quantum.cloud.ibm.com/docs/en/guides/primitive-input-output}{Primitive inputs and outputs}}, which takes circuit depth into consideration.
  \item has\_options – Custom runtime configurations often introduce additional latency, impacting total execution time. For example, gate twirling may be enabled through primitive options\footnote{\href{https://docs.quantum.ibm.com/api/qiskit-ibm-runtime/options-twirling-options}{Primitives TwirlingOptions}}.

\end{itemize}

\subsubsection{Hardware and software used}

The hardware used to complete all the training and benchmarking required for this paper was an Apple MacBook Pro (2021) M1 Max with 32GB of RAM running on macOS 14.5. The same methods could be used in a more powerful computing server without the need of GPUs. In terms of software, our training was completed using Python 3.11\footnote{\href{https://www.python.org/downloads/release/python-3110/}{Python 3.11 Release}}. Among the different libraries utilized we outline pandas \cite{reback2020pandas} and NumPy\cite{harris2020array} for data processing, scikit-learn\cite{scikit-learn} to enable some ML routines, and matplotlib\cite{Hunter:2007} and seaborn\cite{waskom_seaborn:_2021} for data visualization as shown in the Results section.

\subsection{Methods}

\subsubsection{Encoding of the data}

As outlined in the Materials section, we have organized the data columns into various formats for processing. We utilized the \texttt{sklearn.preprocessing} module from the scikit-learn\cite{scikit-learn} library to scale, transform, and prepare the data for model training. Specifically, we employed \texttt{OneHotEncoder}, \texttt{OrdinalEncoder}, \texttt{StandardScaler}, and \texttt{SimpleImputer} to prepare the data for the ML algorithms. Below is a detailed explanation of how each class was used:

a. \texttt{OneHotEncoder}: This encoder is used for categorical column variables, converting them into a binary matrix format suitable for ML algorithms, which enhances prediction accuracy. For each unique category within a feature, \texttt{OneHotEncoder} generates a new binary column, marked as 1 if the category is present and 0 otherwise. This approach is especially beneficial for nominal categories without ordinal relationships.

In our dataset, columns such as primitive\_id, has\_circuits, has\_options, and has\_twirling were encoded using \texttt{OneHotEncoder}. For example, the primitive\_id column contains two distinct values - estimator, and sampler, which when encoded, facilitate more effective training and prediction by the model. Similarly, the other listed columns, which are primarily boolean, benefit from this encoding method.

b. \texttt{OrdinalEncoder}: This encoder transforms categorical features into ordinal integers. It is ideal for categorical variables that inherently possess some order or ranking. The encoder assigns an integer to each category label either based on their alphabetical sequence or a user-defined order. In our dataset, columns such as backend, resilience\_level, and circuit\_type were processed using \texttt{OrdinalEncoder}.

The use of \texttt{OrdinalEncoder} for the backend column was motivated by the impracticality of employing \texttt{OneHotEncoder} due to the extensive variety of backends, which would significantly increase the data dimensions. Experiments indicated that encoding backend with \texttt{OrdinalEncoder} yields effective predictive results. In Qiskit Runtime \cite{Qiskit,javadi2024quantum}, resilience\_level are typically ranked, with lower levels indicating reduced execution time. Given this ordinal nature, \texttt{OrdinalEncoder} was deemed the most suitable encoding method. The circuit\_type column, which includes the values "qpy", "qasm", and "None", also utilizes \texttt{OrdinalEncoder}, although both \texttt{OneHotEncoder} and \texttt{OrdinalEncoder} could be appropriate in this case.

c. \texttt{StandardScaler}: This scaler standardizes features by removing the mean and scaling to unit variance, a common preprocessing requirement for many ML estimators in scikit-learn. It normalizes each feature by subtracting the mean and then dividing by the standard deviation, ensuring that each feature contributes approximately equally to the final prediction. All columns in our dataset were scaled using \texttt{StandardScaler}.

d. \texttt{SimpleImputer}: Handling missing data is a critical challenge in data analysis and ML, addressed through simple imputation methods in the scikit-learn library. This method replaces missing values in the dataset with a predetermined statistic, such as mean, median, mode, or a constant value. In our data preparation process, we used a constant value for imputation. Numerical columns were imputed with \textminus{1}, whereas columns processed with \texttt{OneHotEncoder} and \texttt{OrdinalEncoder} were filled with a placeholder value "NA".

The remaining columns, which are numerical in nature, were used as-is during model training.

\subsubsection{Processing of the data}

During the preprocessing and modeling phases of our study, we used scikit-learn's \texttt{Pipeline} and \texttt{ColumnTransformer} features to expedite our workflows and ensure reproducibility. These components are part of the scikit-learn\cite{scikit-learn} library, which is well-known for providing a comprehensive set of tools for Python ML tasks.

a. Pipeline:

The \texttt{Pipeline}\footnote{\href{https://scikit-learn.org/stable/modules/generated/sklearn.pipeline.Pipeline.html}{sklearn.pipeline.Pipeline}} utility in scikit-learn is essential for encapsulating sequential steps in a data processing and modeling workflow. Each step is represented as a tuple containing a name and an object performing transformation or modeling methods. This encapsulation not only helps to maintain cleaner code, but it also ensures that all processing and model training processes are carried out consistently and in a controlled manner. This is especially important when dealing with cross-validation and model tuning since it prevents data leaks between the training and testing phases and provides a methodical approach to applying transformations and model training to the data.

In our paper, we developed three pipelines for processing diverse input column types, including fillna, encoding, and scaling.

b. ColumnTransformer:

The \texttt{ColumnTransformer}\footnote{\href{https://scikit-learn.org/stable/modules/generated/sklearn.compose.ColumnTransformer.html}{sklearn.compose.ColumnTransformer}} is a powerful tool that allows multiple preprocessing and feature extraction approaches to distinct columns in a dataset. It enables the parallel application of several transformations, which is efficient and effective for datasets including multiple types of data mentioned above. Each transformer in the \texttt{ColumnTransformer} is identified by its name, an estimator or transformation function, and the columns to which it should be applied. This feature is very useful when working with heterogeneous data, since it allows precise modifications to be adapted to the unique properties of each column, improving the dataset's overall prediction performance. In our paper, we have created a \texttt{ColumnTransformer} with the three pipelines mentioned above.

\subsubsection{Training of the model}

Our model training approach is regulated by a carefully crafted configuration file. This configuration file provides the blueprint for our training process, establishing the structural aspects of the model, specifying the X column feature set, Y column, model type to be used, and delineating various hyperparameters required by the learning algorithm.
Hyperparameters include n\_estimators, objective function, alpha, learning\_rate, max\_depth, min\_child\_weight, min\_split\_gain and num\_leaves. The values of these hyperparameters are selected by carrying out different experiments and benchmarking the performance.

Once the data has been processed to the best possible format via pipelines and \texttt{ColumnTransformer}, we proceed with the model training. The entire dataset is split into approximately 94 percent train and 6 percent test data through the selection of a cut-off date. Quantum jobs that completed before or on the cut-off date is used for training and jobs that completed after the cut-off date is used for testing. Next, the model class defined in the configuration file is created, and the processed data is further processed for training. During this phase, the model learns from the data by modifying its parameters to minimize error while following the hyperparameters specified in the configuration file.

\subsubsection{Predicting QPU time using an heuristic method}

Another existing method of predicting QPU time, also developed at IBM Quantum, utilizes a set of predefined formulas to calculate the predicted QPU time for a particular quantum computing job. The predefined formulas are used by leveraging a subset of job metadata similarly as to the ML model. The formulas leverage information including but not limited to the number of shots, primitive options\footnote{\href{https://docs.quantum.ibm.com/api/qiskit-ibm-runtime/options}{Primitive options}}, etc. of the quantum job to derive an estimation for the QPU time. An multiplicative overhead factor is applied in this calculation to adjust for the overhead between executions. The formulas also account for QPU time used for noise learning where applicable.

\section{Results}

In this section we will report the results obtained in the context of this paper to showcase the QPU time prediction for quantum computing jobs. More specifically, we will focus on predicting the QPU time for jobs using sampler/estimator runtime primitives, which are known as the Qiskit primitives\footnote{\href{https://docs.quantum.ibm.com/guides/primitives}{Qiskit primitives}}. 
The sampler runtime primitive is used for circuit sampling, and the estimator runtime primitive is used for expectation value estimation. Together, primitives jobs account for all IBM Quantum jobs. Additionally, for both sampler and estimator jobs, we will compare the prediction results between the ML and heuristic prediction methods.

\subsection{Predicting QPU time}

\begin{figure}
    \centering
    \includegraphics[width=9cm, height=8cm]{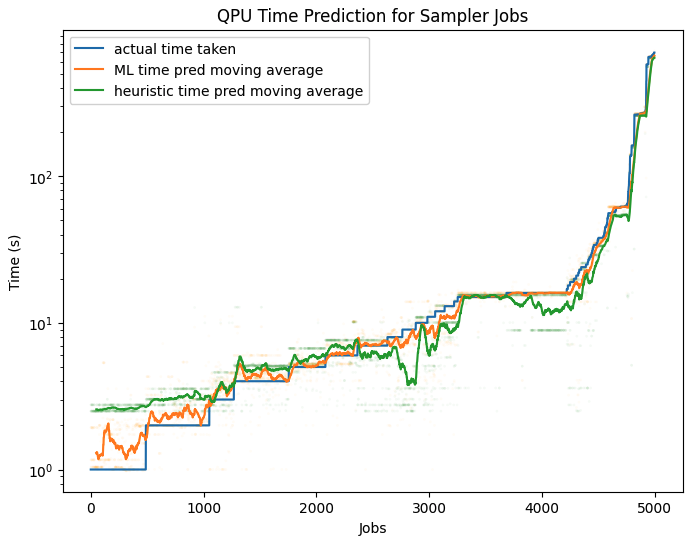}
    \caption{Actual QPU time taken (blue line) vs. ML (orange) and heuristic (green) methods. Orange and green solid lines display moving averages over 50 points while individual predictions are shown as dots.}
    \label{fig:sampler_prediction}
\end{figure}

\begin{figure}
    \centering
    \includegraphics[width=9cm, height=8cm]{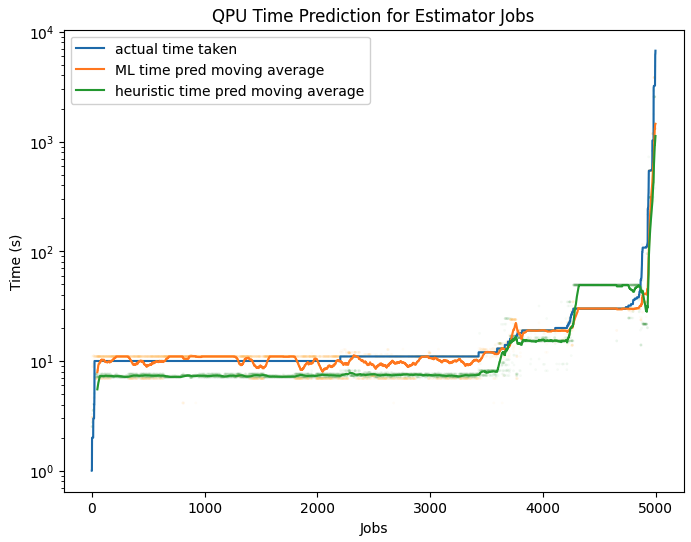}
    \caption{An equivalent set of colors and line formats are used here as in figure \ref{fig:sampler_prediction}.}
    \label{fig:estimator_prediction}
\end{figure}

Figures \ref{fig:sampler_prediction} and \ref{fig:estimator_prediction} both displays a scattered line plot that evaluates the QPU time prediction for sampler and estimator jobs respectively. The x-axis shows the quantum jobs from the test dataset used for QPU time prediction, sorted based on the actual QPU usage from least to greatest. The y-axis represents the QPU time taken/predicted, displayed on a logarithmic scale for readability purposes. The blue line is the actual QPU time taken for each job. The green scattered points represent the predicted QPU time using the heuristic method, and the orange scattered points represent the predicted QPU time using the ML model. The opacity of all scattered points are reduced by setting the alpha field to 0.03 in Matplotlib\footnote{\href{https://matplotlib.org/stable/api/_as_gen/matplotlib.pyplot.scatter.html}{matplotlib.pyplot.scatter}} to improve the graph's readability. To further enhance readability, a line representing the moving average is added for each set of scattered points. 

The green line is the moving average for the green scattered points and the orange line represents the moving average for the orange scattered points. The moving average is calculated using numpy's \texttt{convolve} method\footnote{\href{https://numpy.org/doc/stable/reference/generated/numpy.convolve.html}{numpy.convolve}} with window size of 50. In the scenario of a perfect prediction, the scattered point would sit directly on top of the solid blue line (actual time taken). An overestimation occurs when a particular scattered point lies above the blue line, signifying that the predicted QPU time is greater than its actual QPU time taken. Similarly, underestimation occurs when a particular scattered point is plotted beneath the blue line. The absolute error of the prediction increases as the vertical distance increases from a particular scattered point to the blue line.

For both figures \ref{fig:sampler_prediction} and \ref{fig:estimator_prediction}, it can be observed that the orange line is closer to the blue line compared to the green line. This signifies that the prediction provided by the ML model is generally more accurate than the heuristic method for both sampler and estimator jobs.

In figure \ref{fig:sampler_prediction} for sampler jobs, the predictions using the heuristic method had a relatively larger overestimation for the first approximately 1000 jobs, where the actual QPU time taken was shorter. For the same set of quantum jobs, the ML model also overestimated the QPU time, however, the degree of overestimation was much smaller in comparison. Time predictions by the heuristic method were also visibly underestimating jobs around 2500 to 3200 and jobs around 3800 to 4200. The predictions using ML were significantly better for these job ranges as shown by the orange line being closer to the blue line in these areas.

As displayed in figure \ref{fig:estimator_prediction} for estimator jobs, the predictions using the heuristic method were generally underestimated for the first approximately 4000 jobs. The underestimations are then followed by a spike in overestimation for jobs around 4200 to 4800. On the other hand, the prediction using the ML model performed significantly better in the above areas, the orange line was generally close to the blue line throughout the graph.

A major strength of using ML over the heuristic method is that the ML model learns about the hardware capabilities from its extensive set of training data. Therefore, the ML model is capable of providing QPU time predictions based on the hardware capabilities of the associated IBM Quantum backend. Our heuristic methods currently do not employ hardware profiles of specific IBM Quantum backends. Consequently, the heuristic predictions do not consider the hardware capability of the quantum backend used for a particular job.

\subsection{QPU Time Prediction Prediction Accuracy}

\begin{figure}
    \centering
    \includegraphics[width=9cm, height=8cm]{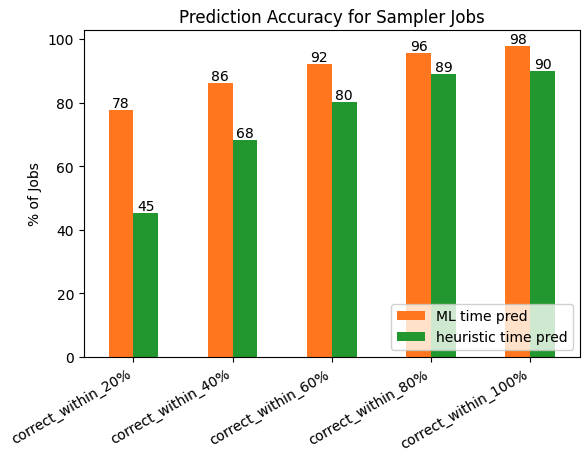}
    \caption{QPU time prediction accuracy for sampler jobs using ML versus heuristic prediction methods.}
    \label{fig:sampler_correctness}
\end{figure}

\begin{figure}
    \centering
    \includegraphics[width=9cm, height=8cm]{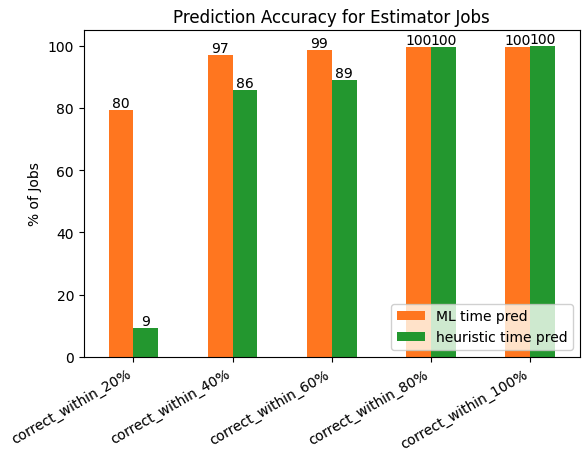}
    \caption{QPU time prediction accuracy for estimator jobs using ML versus heuristic prediction methods.}
    \label{fig:estimator_correctness}
\end{figure}

Figures \ref{fig:sampler_correctness} and \ref{fig:estimator_correctness} evaluates the predictive accuracy for sampler and estimator jobs respectively. In both figures, each bar represents the percentage of quantum jobs with predictions that satisfy the associated level of accuracy. More specifically, quantum jobs represented by the orange bars are predicted using the ML model whereas the green bars represent quantum jobs with predictions using the heuristic method.

The x-axis shows 5 categories: correct\_within\_20\%, correct\_within\_40\%, correct\_within\_60\%, correct\_within\_80\%, and correct\_within\_100\% respectively. In order to derive the data shown in the figures, the percentage error is calculated based on the formula

\[
\left| \frac{\text{time\_predicted} - \text{time\_taken}}{\text{time\_taken}} \right| \times 100\%
\]For example, the correct\_within\_20\% category will include job predictions with percent error of less or equal to 20\%. In other words, the prediction was within 20\% of the actual time taken.

The y-axis represents the percentage of jobs that satisfies each category. For example, in figure \ref{fig:sampler_correctness}, 78\% of the sampler jobs had percentage error within 20\% using prediction from the ML model. The same value lowers to 45\% when using heuristic methods.

As shown in figures \ref{fig:sampler_correctness} and \ref{fig:estimator_correctness}, the predictive accuracy is higher using the ML model for both sampler and estimator jobs. The percentage of jobs that satisfies each category is higher for predictions using ML with the exception that the categories correct\_within\_80\% and correct\_within\_100\% in figure 4 is 100\% for predictions using both ML and the heuristic method. The difference between the two bars is the greatest for the category correct\_within\_20\% and the gap generally tightens as the percent error increases. For example, in the correct\_within\_20\% category, the difference between the two bars is 33 percentage points for sampler jobs and 71 percentage points for estimator jobs. As the percentage error increases to 100\% as shown by the category correct\_within\_100\%, the difference lowers to 8 percentage points and 0 percentage points for sampler and estimator jobs respectively.

The percentage of estimator jobs with predictions using the heuristic method is 9\% for the correct\_within\_20\% category, significantly lower than the value for rest of the categories. This may be due to the more sophisticated nature of estimator, causing estimator jobs to be more frequently split into more batches. Therefore, it becomes challenging for the heuristic method to provide predictions within 20\% of correctness without backend-specific information. On the other hand, the predictions using our ML model can predict 80\% of estimator jobs within 20\% of percent error, since our model considers much more data fields per estimator job. As mentioned in the subsection above, the model is also able to learn about the associated IBM Quantum backend from its training data.

\subsection{Calculation of the multiplicative safety factor}

\begin{figure}
    \centering
    \includegraphics[width=9cm, height=8cm]{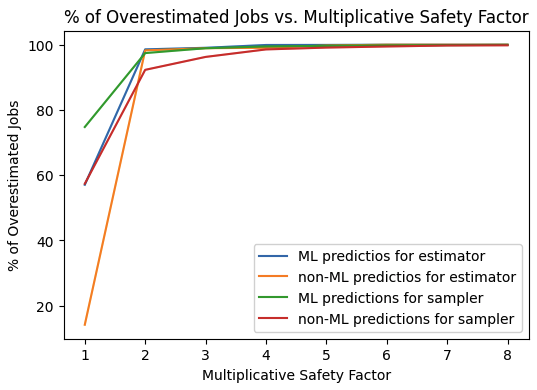}
    \caption{Percentage of overestimated jobs after applying a particular multiplicative safety factor to its original prediction.}
    \label{fig:safety_factor}
\end{figure}

A multiplicative safety factor may be applied to the prediction result to mitigate underestimation. The safety factor is a deliberate overestimation applied to the predicted times to ensure a margin of error. The application of a multiplicative safety factor may be especially helpful in scenarios where underestimations may cause disappointment or even alarm, as the job is running longer than expected.

Figure \ref{fig:safety_factor} depicts the relationship between the applied multiplicative safety factor and the percentage of overestimated sampler and estimator jobs, using both the ML and heuristic prediction methods. The x-axis represents the applied multiplicative safety factor, ranging from 1 to 8. The y-axis quantifies the percentage of jobs for which the actual time taken was less than the predicted time after applying the safety factor, indicating an overestimation. For example, a value of 100 on the y-axis would imply that 100\% of the jobs were overestimated once applying the associated x-axis safety factor to the original prediction.

Based on figure \ref{fig:safety_factor}, all jobs approach near 100\% overestimation after applying a safety factor between 2 to 3, with the exception of sampler jobs using heuristic methods, which reaches close to 100\% once the safety factor increases to 4. 

These visualizations are critical for evaluating the impacts of the safety factor on prediction accuracy across different runtime primitives and prediction methods. The process of applying safety factors requires various tailoring to match the characteristics of jobs for each primitive, rather than applying a uniform factor across the diverse computational tasks.

\section{Discussion}

This paper presents a novel contribution in predicting the QPU time of quantum jobs using ML techniques, which can be useful to enhance the current quantum computing stacks. For instance, predicting QPU time could be used to improve the scheduling of quantum computing jobs execution or to inform users about how long their jobs are expected to run on a quantum computer.

As shown in this paper, QPU time predictions using ML methods can be significantly more accurate compared to heuristic method for both sampler and estimator jobs. By leveraging the extensive set of training data, the ML model is able to learn about the performance of IBM Quantum backends, and tailor the prediction based on the backend used. However, predicting the QPU time is not a straightforward process, even when using tools such as ML algorithms. This is due to the variability of quantum jobs executed and their related options on an already evolving environment like actual quantum computing platforms, such as IBM Quantum's. In this paper, we have demonstrated a workflow for building ML prediction models that could in the future be automated to track changes in execution time characteristics. More importantly, these updates can be done without expert effort to tune a heuristic model. With the application of safeguards, such as safety factors, the ML QPU time prediction model may be employed to enhance existing quantum computing systems. The predicted values could be used as a threshold for stopping stuck operations or scheduling operations between different users, among other examples.

The work initiated in this project lays the groundwork for incorporating new intelligent features in quantum computing systems. We have identified several other potential tasks where ML could be beneficial, including: compilation time, hardware resources required to process quantum jobs (with or without time constraints), assessing the validity of a job (determining if a job will run or fail on a quantum computing platform), among others. Our plan is continue working on them in future research works.

\section*{Acknowledgments}
The authors acknowledge feedback and insightful discussions with Blake R. Johnson.

\bibliographystyle{IEEEtran}
\bibliography{references}
\end{document}